\documentclass[sigconf,nonacm]{acmart}
\AtBeginDocument{%
  }

\begin{document}

\title{GitFarm: Git as a Service for Large-Scale Monorepos}

\author{Preetam Dwivedi}
\orcid{0009-0005-0778-1385}
\affiliation{%
  \institution{Uber Technologies Inc.}
  \country{USA}
}
\email{preetam@uber.com}

\author{Akshay Hacholli}
\orcid{0009-0006-8228-4618}
\affiliation{%
  \institution{Uber Technologies Inc.}
  \country{USA}
}
\email{hacholli@uber.com}

\author{Adam Bettigole}
\orcid{0009-0001-4052-4242}
\affiliation{%
  \institution{Uber Technologies Inc.}
  \country{USA}
}
\email{abettigole@uber.com}

\renewcommand{\shortauthors}{Dwivedi et al.}

\begin{abstract}
At the scale of Uber’s monorepos, traditional Git workflows become a fundamental bottleneck. Cloning multi-gigabyte repositories, maintaining local checkouts, periodically syncing from upstream, and executing repetitive fetch or push operations consume substantial compute and I/O across hundreds of automation systems. Although CI (Continuous Integration) systems such as Jenkins and Buildkite provide caching mechanisms to reduce clone times, in practice, these approaches incur significant infrastructure overhead, manual maintenance, inconsistent cache hit rates, and cold start latencies of several minutes for large monorepos. Moreover, thousands of independent clone and fetch operations add heavy load on upstream Git servers, making them slow and difficult to scale.

To address these limitations, we present \textbf{GitFarm}, a platform that provides \textit{Git as a stateful, identity-scoped, repository-centric execution service} through a gRPC API. GitFarm decouples repository management from clients by executing Git operations remotely within secure, ephemeral sandboxes backed by pre-warmed repositories. The system enforces identity-scoped authorization, supports multi-command workflows, and leverages specialized backend clusters for workload isolation.

For clients, this design eliminates local clones, provides a ready-to-use checkout in less than a second, and significantly lowers client-side compute and I/O overhead by offloading operations to GitFarm. Also, client services no longer experience cold starts (up to 15 minutes) due to initial clones of the monorepos on each host. The results demonstrate that Git as a service provides substantial performance and cost benefits, while preserving the flexibility of native Git semantics.
\end{abstract}

\maketitle

\section{Introduction}
Source control systems form the backbone of modern software engineering. Git \cite{git2025} is the most widely used distributed version control system, providing a decentralized model for managing source code history and collaboration. As organizations scale to thousands of engineers and millions of source files, the performance and reliability of version-control operations become critical to developer productivity. Large repositories (monorepos) \cite{brito2018monorepos} have emerged as a dominant model at companies such as Google \cite{potvin2016googlemonorepo}, Meta \cite{facebook2019mononoke,facebook2022sapling}, and Microsoft \cite{microsoft2018vfsforgit,microsoft2020scalar}, enabling unified dependency management and global refactoring. Yet these advantages come with steep infrastructure costs, as conventional Git workflows were not designed for repositories containing millions of commits and gigabytes of history. Low et al. \cite{low2023gitisfordata} highlighted similar scalability limits in their adaptation of Git for terabyte-scale datasets.

At Uber, various automation systems and developer services invoke Git millions of times per day across multiple monorepos spanning Go, Java, Python, Web, Android, and iOS. Historically, each system maintained its own full repository checkout, even for lightweight operations such as reading up-to-date files. This consumed large amounts of compute, memory, and storage, and, more importantly, generated sustained load on upstream Git servers, as every system performed its own clone and fetch operations, hampering the performance of the system and developer workflows.

Traditional caching mechanisms offer only limited relief. CI platforms such as Jenkins \cite{jenkinsgitplugin2023reference} and Buildkite \cite{buildkite2023caching,buildkite2020issue734} reuse local mirrors to speed up clones, yet per-agent caches quickly diverge and require expensive fetches on CI job execution. Even with caching, large monorepos often spend several minutes preparing before execution. These solutions accelerate CI jobs but do not scale to the continuous, multi-service workflows GitFarm targets.

To address these limitations, we built GitFarm, a platform that provides Git as a Service through a gRPC API. Instead of cloning repositories locally, clients issue remote procedure calls  \cite{google2020reapi} to GitFarm, which executes Git operations inside secure, ephemeral sandbox containers  \cite{gvisor2018hotcloud} provisioned from pre-warmed pools of repositories. This design decouples repository management from individual clients, centralizes synchronization with upstream servers, and delivers consistent low-latency performance. GitFarm supports both read and write operations, enforces identity-scoped authorization, enables multi-command workflows in a single connection, and isolates workloads through specialized backend clusters.

GitFarm has been rolled out across Uber’s developer platform since early 2025, enabling use cases such as code-review validation and raising audit pull requests. In production, it reduces the client service initialization latency from up to 15 minutes to less than a minute, and lowers the client-side resource utilization by more than 70\%. By eliminating redundant clones and centralizing repository synchronization, GitFarm reduces upstream Git traffic, improves overall system reliability, and frees resources for other critical operations.

This paper makes the following contributions to the design of scalable
version-control infrastructure for large monorepos:

(1) We introduce \textit{repository-centric execution}, a model in which
Git operations execute against centrally-managed, pre-warmed repositories
rather than per-client clones.

(2) We present \textit{stateful Git execution sessions}, enabling
output-dependent, multi-command Git workflows with consistency guarantees not
supported by traditional CI systems or Git hosting APIs.

(3) We design \textit{identity-scoped, sandboxed Git execution}, decoupling
authorization from client environments while preserving native Git semantics.

(4) Through production deployments at Uber, we demonstrate that this model
substantially reduces latency, resource usage, and operational complexity across
a range of Git-intensive automation workloads.

We describe our motivation and the limitations of existing Git infrastructures,
detail the design and implementation of GitFarm’s architecture, and evaluate its
impact on performance and resource efficiency in production at Uber.

\section{Background}
\begin{figure*}[!t]
    \centering
    \includegraphics[width=1\textwidth]{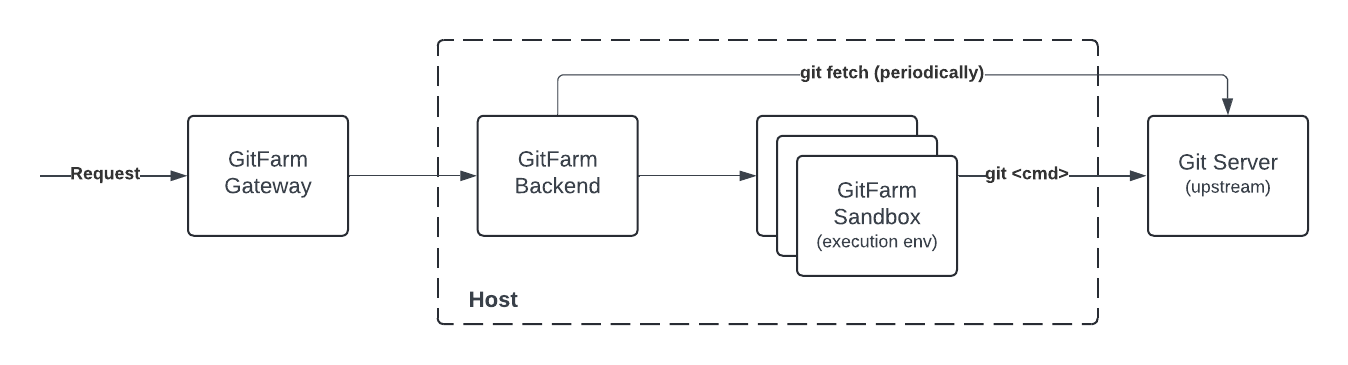}
    \caption{GitFarm High Level Architecture}
    \Description{GitFarm High Level Architecture}
    \label{fig:gitfarm-architecture}
\end{figure*}

As Uber’s codebase continues to grow, so does the complexity of managing Git repositories across a large and diverse set of automation systems. Many internal services rely on Git operations for observing repository changes, checking merge conflicts, rebasing branches, and pushing derived references. At scale, performing these operations efficiently and reliably becomes increasingly challenging.

\subsection{Managing Local Checkouts for Monorepos}

Historically, each system at Uber that required access to Git repositories maintained its own local checkout. This approach is both time-consuming and resource-intensive, particularly for checkouts of large monorepos. For example, cloning Uber’s Go monorepo can take approximately 15 minutes and requires roughly 4 CPU cores, 16 GB of memory, and more than 40 GB of disk space. A single service node that must clone and synchronize multiple major Uber monorepos (e.g., Go, Java, Web, Python, Android, and iOS) can easily require 12 CPU cores, 48 GB of memory, and 64 GB of disk space.

Executing a \textit{git clone} requires the client to download all reachable repository objects \cite{git2024packobjects} and write them into packfiles under \texttt{.git/objects/pack} \cite{chacon2024progitpackfiles}, generating heavy sequential disk I/O. Git compresses objects using \textit{zlib} and applies delta compression between related objects; resolving these deltas requires decompression and in-memory reconstruction, making the process CPU- and memory-intensive \cite{github2021monorepomaintenance}. Creating the working tree further stresses the file system by writing thousands of files to disk. Although incremental fetches and pushes are less expensive, clients must still retain sufficient resources to maintain large local checkouts and periodically synchronize them to remain up to date.

Techniques such as shallow clones, partial clones \cite{git2024partialclone}, or single-branch clones can reduce initial data transfer, but they do not fundamentally address scalability. Each request still requires the upstream Git server to enumerate objects, generate packfiles, and stream compressed data to the client. As the number of clients grows, this repeated work increases server-side CPU and I/O load, and becomes a bottleneck. Moreover, these optimizations impose functional limitations: operations such as \textit{merge-base} or \textit{bisect} may fail on shallow clones, and partial clones often break tooling assumptions that are common in large monorepos.

\subsection{CI Platforms as a Git Execution Mechanism}

To avoid managing long-lived local checkouts directly, many services rely on CI platforms to execute Git operations. CI (Continuous Integration) systems provide standardized execution environments, isolation, and scheduling, making them a convenient integration point for automation \cite{jenkins2024distributedbuilds,buildkite2024agentv3}.

In these systems, Git operations are performed as part of CI jobs that allocate an agent, prepare a workspace, and synchronize the repository before executing any logic. While this model works well for build and test workflows, it introduces significant overhead for lightweight Git operations. Repository checkout and workspace preparation occur on the critical path of every job, often dominating execution time for automated tasks that do not involve compilation or testing.

At Uber’s scale, this approach leads to substantial agent-side latency and inefficient resource usage. As a result, simple Git operations, such as inspecting a commit, computing a merge base, or pushing a git ref, are forced through heavyweight CI workflows, incurring unnecessary scheduling delays, network transfers, and repository checkout initialization overhead.

These challenges highlight a fundamental mismatch between CI-oriented execution models and the needs of Git-oriented automated services. As repositories grow and automated workloads proliferate, relying on per-client local checkouts or CI jobs for Git access becomes increasingly inefficient and makes integration significantly more complex (especially, for workloads that need a back and forth communication).

\section{Architecture}

When a request is submitted to GitFarm, it is first received by the GitFarm Gateway, which forwards the request to the GitFarm Backend. Within the Backend, an available node is chosen to process the request. The request is then executed in a sandbox on that node, where a pre-warmed repository checkout is mounted, as shown in Figure \ref{fig:gitfarm-architecture}.

\subsection{Gateway - Entrypoint}

The Gateway is responsible for authenticating and authorizing incoming requests and routing them to the appropriate backend. Upon receiving a request, the Gateway identifies the client and verifies that the client has permission to access the requested repository. Requests from clients lacking the required privileges are denied.

The Gateway also functions as a load balancer for the GitFarm backends. It continuously tracks backend node availability by monitoring periodic heartbeats and status updates that report the number of available sandboxes on each node for each repository. This state is maintained in a Redis data store. For each incoming request, the Gateway selects the backend node with the highest number of available repository checkouts for the requested repository, and marks the selected repository checkout as occupied. It then releases the sandbox upon completion by updating the Redis state. If no backend nodes are available for a given repository, the Gateway rejects incoming requests, effectively throttling the workload due to insufficient resources.

\subsection{Backend - Request Processor}

The GitFarm Backend forms the core execution layer of the system. It maintains a warm, up-to-date repository state by periodically synchronizing on-disk repositories with their upstream remotes (via \texttt{git fetch}). The Backend manages a pool of isolated execution environments (sandboxes) used to execute Git commands issued through API requests.

\begin{figure*}[!t]
    \centering
    \includegraphics[width=\textwidth]{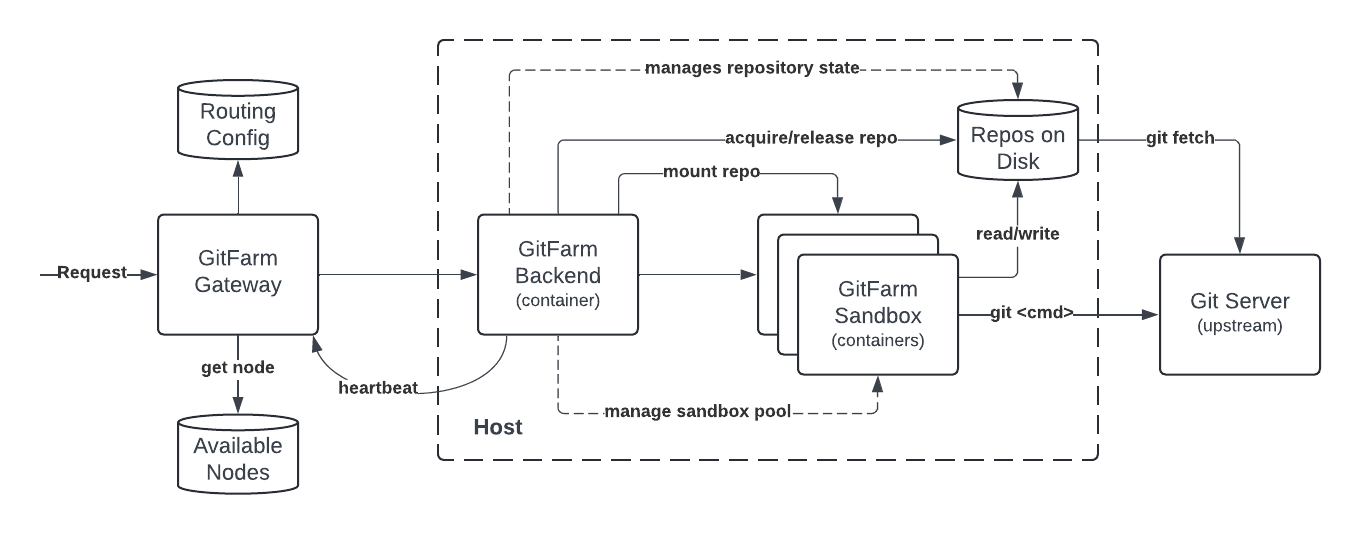}
    \caption{GitFarm Backend Architecture}
    \Description{GitFarm Backend Architecture}
    \label{fig:gitfarm-backend}
\end{figure*}

For each incoming request, the Backend provisions the appropriate repository state, identity, and execution context, and mounts them into a sandbox to perform the requested operation, ensuring isolation and consistency across executions, as illustrated in Figure \ref{fig:gitfarm-backend}.

\subsubsection{Maintaining Repository States}

Each GitFarm Backend node maintains a single on-disk bare clone for each repository. These clones are kept up to date through an event-driven synchronization strategy (push-based updates) that fetches changes as soon as they become available. We also perform periodic synchronizations with the upstream repository by executing \texttt{git fetch} every five minutes, to ensure the repository state remains fresh in the event of missed or delayed events.

\paragraph{Freshness and Staleness Semantics.}
GitFarm does not enforce a single global freshness guarantee across all workloads. Instead, it exposes repository state that is eventually consistent with the upstream repository, while allowing clients to explicitly control freshness on a per-request basis.

For workloads that require the most up-to-date repository state, clients may invoke an explicit \texttt{git fetch} as part of their execution session prior to running any other Git commands. This ensures that all subsequent operations within the session observe the latest available commits from the upstream repository. Other workloads that tolerate bounded staleness may rely solely on the backend’s repository synchronization mechanisms to minimize execution latency and avoid redundant fetches. 

This design allows GitFarm to accommodate a diverse range of workloads and provides maximum flexibility to clients.

\subsubsection{Sandbox - Execution Environment}

A sandbox is an ephemeral execution environment (container) in which all Git operations for a given request are executed. Each request runs in complete isolation within its own sandbox and uses a dedicated Git checkout, ensuring separation from other concurrent requests in GitFarm. This isolation model is similar to container-based execution environments such as gVisor \cite{gvisor2018hotcloud}.

Sandboxes are pre-initialized to reduce request latency and are protected by strict security boundaries. Access privileges within a sandbox are scoped to the caller’s identity.

\begin{figure}[htbp]
    \centering
    \includegraphics[width=1\linewidth]{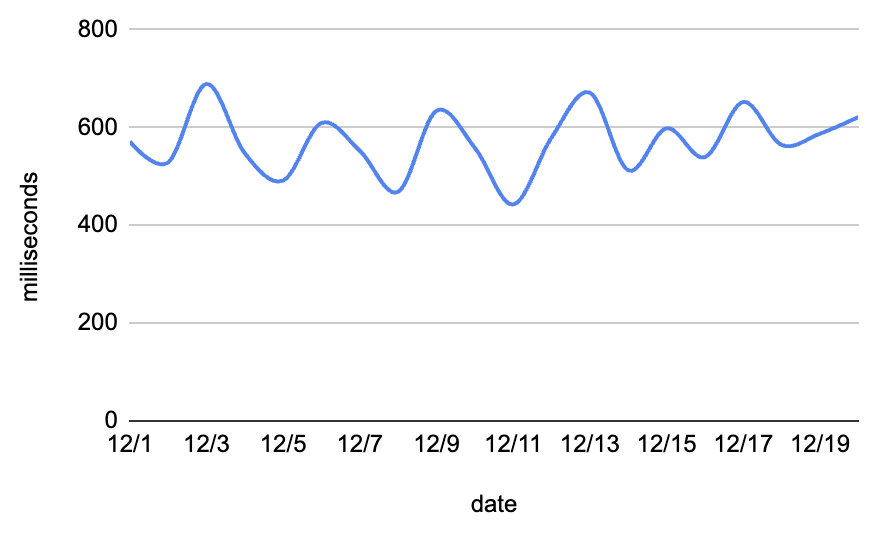}
    \caption{P95 Latency to Acquire Sandbox}
    \Description{P95 Latency to Acquire Sandbox}
    \label{fig:acquire-sandbox-latency}
\end{figure}

\subsubsection{Pooling – Eliminating Cold Start}

Creating a new repository checkout and provisioning a sandbox container on demand for each request is highly resource-intensive. Spinning up a sandbox container typically takes 1–2 seconds, and materializing a repository checkout from a local bare clone (periodically synchronized from upstream) can take up to 3 minutes. Such per-request latencies are unacceptable and do not scale for high-volume workloads.

To address this, GitFarm employs a pooling model for both repository checkouts and sandbox containers, following a design pattern commonly used in distributed compilation and remote execution systems to amortize setup costs across requests \cite{matev2019fastdist}. The Backend maintains a fixed-size pool of repository checkouts on disk, each synchronized from a local bare clone of the repository. These pre-warmed checkouts are immediately available for request execution.

Similarly, the Backend pre-creates a fixed-size pool of sandbox containers, each initialized with an isolated execution environment and a dedicated mount point. When a request arrives, the Backend acquires an available sandbox from the pool and mounts a repository checkout from the repository pool into the sandbox. This design eliminates on-demand provisioning and allows multiple repositories to be efficiently served within a single cluster.

With pooling in place, the overhead of providing a sandbox with a ready-to-use repository checkout is reduced to less than a second (as seen in Figure \ref{fig:acquire-sandbox-latency}), substantially lowering request latency. This optimization enables the system to dedicate resources to executing Git operations rather than incurring repeated initialization costs associated with cold starts.

\subsection{Stateful Request Chaining}
Many automation workflows require executing multiple Git operations sequentially, where the output of one command is consumed by subsequent commands, all operating within the same repository checkout. For example, computing the merge base between two branches and publishing it under a derived reference requires capturing the output of \texttt{git merge-base} and using it to push a new Git ref.

GitFarm supports such workflows through a bidirectional gRPC streaming API that allows clients to execute a sequence of Git commands within a single persistent session, with full access to stdin and stdout. This model preserves a consistent repository checkout across commands while enabling output-dependent command chaining, minimizing connection setup and environment initialization overhead.

\subsection{Clustering}
\begin{figure}[htbp]
    \centering
    \includegraphics[width=1\linewidth]{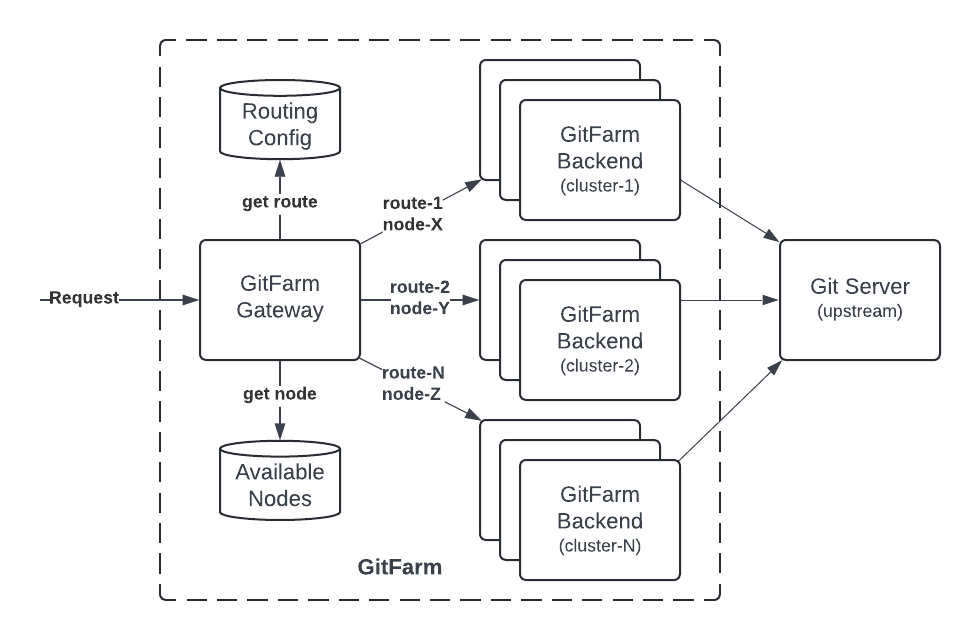}
    \caption{GitFarm Backend Clustering}
    \Description{GitFarm Backend Clustering}
    \label{fig:gitfarm-cluster}
\end{figure}

Clustering in GitFarm refers to deploying multiple GitFarm Backend nodes grouped into logical clusters, each purpose-built to serve a specific, uniform use case as illustrated in Figure \ref{fig:gitfarm-cluster}. This model ensures predictable performance characteristics while avoiding “noisy neighbor” effects caused by heterogeneous workloads sharing the same backend nodes.

At Uber, we operate multiple specialized clusters in addition to a generic shared cluster. Specialized clusters are tailored for high-throughput or latency-sensitive workloads with well-defined access patterns, whereas the shared cluster supports lighter-weight use cases and provides a fast, low-friction integration path for new clients, while isolating heterogeneous workloads
using cluster-level resource partitioning strategies similar to those explored in distributed compilation systems \cite{distcom2021}.

When a new use case arises, we conduct a review to determine whether the workload can be safely accommodated by an existing cluster or requires a dedicated one. For use cases that warrant isolation, we perform sizing analysis and apply cluster-specific configuration updates (e.g., resource limits, sandbox pool sizing, synchronization policies) prior to onboarding.

Routing decisions are enforced by the GitFarm Gateway, which selects the appropriate cluster for each incoming request based on centrally-managed placement policies. These policies associate each client with a designated cluster, enabling fast and deterministic request routing while preserving workload isolation. This design allows GitFarm to support specialized clusters alongside a shared cluster, without exposing complexity to clients.

\subsection{API Specification}

GitFarm exposes an execution interface (Figure \ref{fig:gitfarm-exec-api}) that enables clients to execute a sequence of Git commands within a single repository checkout. The interface is designed to support multi-step workflows while preserving repository consistency and minimizing execution overhead.

\begin{figure}[htbp]
\begin{verbatim}
Exec(repo_id, workspace_type, commands)→CommandResults

Commands: list<Command>
  Command:
    alias: string
    binary: string
    arguments: list<string>
    stdin: string (optional)
    environment: map<string, string>

CommandResults: list<CommandResult>
  CommandResult:
    alias: string
    exit_code: int
    stdout: string
    stderr: string
\end{verbatim}
\caption{Pseudocode interface for the GitFarm execution API}
\Description{Pseudocode interface for the GitFarm execution API}
\label{fig:gitfarm-exec-api}
\end{figure}

\paragraph{Execution Semantics.}
Each \texttt{Exec} invocation establishes a persistent session bound to a single logical repository checkout. Commands are executed sequentially in the order they are received, and the result of each command is returned to the client as a corresponding \texttt{CommandResult}. The \texttt{alias} field is used to correlate command invocations with their
outputs, but they're also returned in the same order in which they were executed.

\paragraph{Guarantees.}
GitFarm provides the following guarantees for each execution session:  
\begin{itemize}
    \item \emph{Isolation}: commands execute within an isolated sandbox environment;
    \item \emph{Consistency}: all commands observe a consistent repository state throughout the session;
    \item \emph{Determinism}: command execution order matches the order in which they're defined in the request.
\end{itemize}

\paragraph{Error Handling.}
Command failures are surfaced via non-zero exit codes and populated standard error output. Fatal errors terminate the session.

\section{Evaluation}

\subsection{Setup}
We conducted our evaluation on a shared GitFarm cluster in production consisting of four backend nodes. Each backend node runs on a Linux host provisioned with 48 CPU cores, 250~GB of memory, and 1.5~TB of local SSD storage.

Each production backend node maintains a bare clone of every supported monorepo, periodically synchronized with upstream Git servers, along with fixed-size pools of working checkouts derived from these clones. In our setup, each node maintains a pool of 100 sandbox containers for request execution, as well as 30 checkouts for the Go monorepo (approximately 900~GB of disk space), and 10 checkouts each for the Java, Android, Web, Python, and iOS monorepos (approximately 110~GB per repository).

Repository and sandbox pool sizes are determined empirically by measuring the CPU overhead of repository synchronization, general system maintenance, and sandbox management. Pools are sized to keep this overhead under 20\% of node CPU capacity, reserving the remaining headroom for request execution.

We evaluate GitFarm using the following production-inspired workloads: 
\begin{enumerate}
    \item compliance auditing for bypassed development workflows,
    \item automatically changing pull request base branches,
    \item read-only inspection of up-to-date code ownership files.
\end{enumerate}

Together, these workloads demonstrate GitFarm’s impact on resource efficiency, execution latency, and workflow simplicity.

\subsection{Case Studies}

\subsubsection{Case Study 1: Compliance Auditing for Bypassed Development Workflows}

Many production environments enforce strict compliance requirements on commits prior to merge and deployment, such as mandatory builds, test execution, code reviews, and approvals. In exceptional circumstances, these safeguards may be bypassed, requiring post-merge and post-deploy auditing to ensure accountability and retrospective review.

\paragraph{Setup.}
We evaluate GitFarm using a production service that audits commits which bypass standard compliance checks. The service processes approximately 10k–20k deployment and Git push events per hour across about 9,000 repositories. For each event, it determines whether any commits bypassed the review process and, if so, generates an audit by publishing a pull request for the unreviewed commits. This requires fetching the relevant base and head references from the original push or deployment event and pushing them to corresponding remote branches.

Originally, the service relied on Buildkite to perform the Git operations, as maintaining local checkouts for thousands of repositories is highly resource-intensive and unnecessarily couples repository management and synchronization logic with the service’s business logic.

\paragraph{Metrics.}
We report end-to-end execution latency measured at the client and focus on median (p50) values. The latency captures the time required to compute the relevant base and head references and push the resulting refs to the remote repository in preparation for pull request creation; the time to create the pull request itself is excluded. We do not report tail latencies (p99) for this workload because request volume is low at times, and tail behavior is dominated by variability in Git operations, rather than by the execution infrastructure itself.

\begin{figure}[htbp]
    \centering
    \includegraphics[width=1\linewidth]{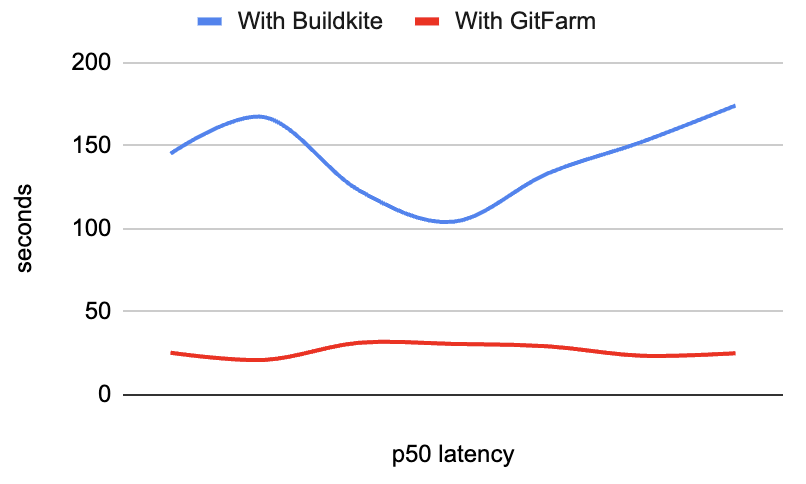}
    \caption{P50 Execution Latency for Compliance Auditing over A Week}
    \Description{P50 Execution Latency for Compliance Auditing over A Week}
    \label{fig:code-compliance-latency}
\end{figure}

\paragraph{Baseline (Buildkite).}
When using Buildkite, each audit invocation provisioned a Buildkite worker, initialized a repository checkout, fetched the unreviewed references, and pushed the corresponding audit branches. Since every invocation required full workspace setup and repository synchronization, execution latency was dominated by Buildkite environment initialization overhead, resulting in a p50 latency of approximately 110--160 seconds over a week (Figure \ref{fig:code-compliance-latency}).

\paragraph{GitFarm-Based Execution.}
After onboarding to GitFarm, the service replaced Buildkite managed repository operations with API calls that execute the required Git commands against centrally maintained repositories. The audit workflow now runs as a short-lived GitFarm execution session, still with no local repository state maintained by the client. As a result, end-to-end p50 latency was reduced to approximately 20--30 seconds (Figure \ref{fig:code-compliance-latency}), representing more than an 80\% reduction.

\paragraph{Discussion.}
By removing Buildkite environment provisioning and local repository management from the critical path, GitFarm reduces execution latency to the cost of the underlying Git operations. This case study demonstrates that GitFarm significantly lowers latency for automated workflows and avoids the operational complexity of Buildkite-based execution.

\subsubsection{Case Study 2: Changing pull request base branches}

Our second evaluation examines an automated workflow for updating the base branch of a pull request. The workflow fetches the relevant branches, computes the merge base, and publishes the resulting commit under a remote reference, which is used as the new base. Using GitFarm, these operations execute as a single session that runs a short sequence of Git commands (\texttt{fetch}, \texttt{merge-base}, \texttt{rev-parse}, and \texttt{push}) within the same sandbox backed by a warm repository, illustrating lightweight, stateful Git interactions with output-dependent request chaining.

\paragraph{Setup.}
This workload is deployed in production and processes low-volume traffic at approximately 1 request per second. As a newly introduced use case, no prior CI or local checkout-based implementation exists for direct comparison.

\paragraph{Metrics.}
We report client-observed end-to-end execution latency and focus on median (p50) values. Given the low request volume, tail latencies are not representative of steady-state performance and are therefore omitted.

\begin{figure}[htbp]
    \centering
    \includegraphics[width=1\linewidth]{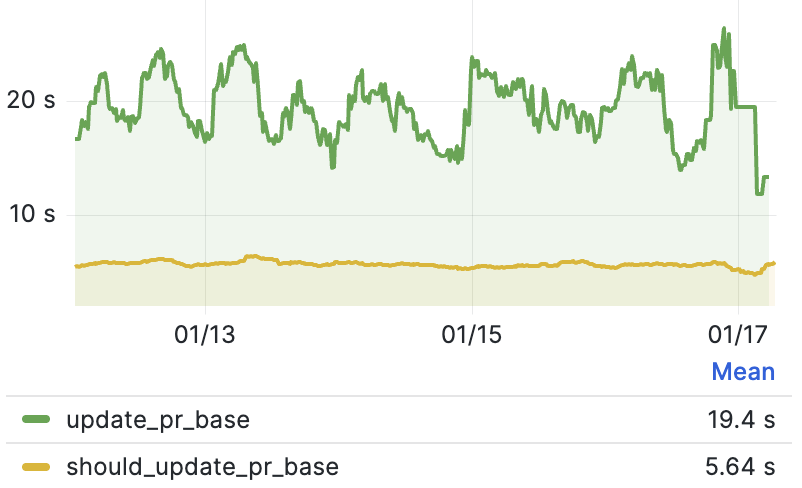}
    \caption{P50 Workflow Latency}
    \Description{P50 Workflow Latency}
    \label{fig:code-review-latency}
\end{figure}

\paragraph{Latency.}
The workflow consists of two phases. The first phase checks whether the base of the pull request (PR) needs to be updated, and if so, the second phase performs the update (which involves a fetch and a push). As shown in Figure \ref{fig:code-review-latency}, the end-to-end workflow exhibits a p50 latency of approximately 25 seconds for completing both phases. 

\begin{figure}[htbp]
    \centering
    \includegraphics[width=1\linewidth]{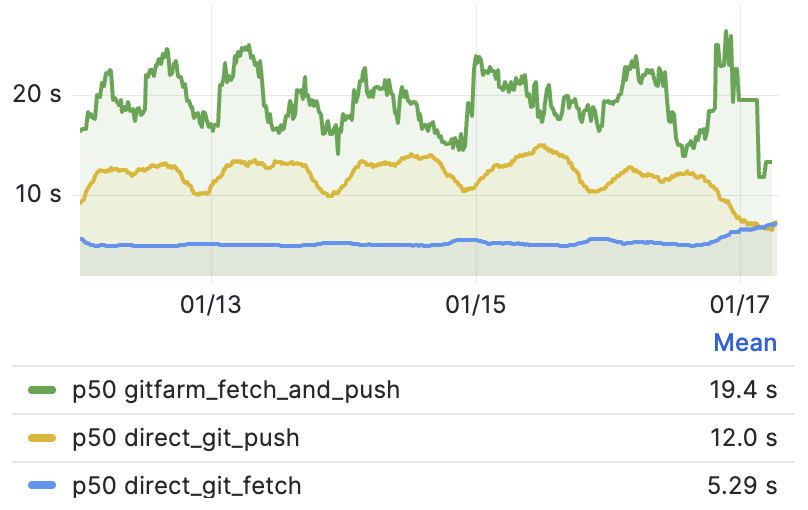}
    \caption{P50 Git Fetch and Push Latency comparison}
    \Description{P50 Git Fetch and Push Latency comparison}
    \label{fig:latency-comparision}
\end{figure}

This latency is primarily driven by Git operations, specifically fetch (p50 of 5s) and push (p50 of 12s), rather than by execution overhead from GitFarm itself. In large monorepos, push operations are inherently expensive due to server-side validation, object processing, and reference updates. This behavior is further illustrated by the p50 latency comparison between GitFarm and direct git server interactions in Figure \ref{fig:latency-comparision}, which shows comparable costs dominated by these underlying Git operations.

\paragraph{Discussion.}
By eliminating local repository checkouts, the critical path for this workflow is reduced to executing Git commands against centrally maintained repositories. Although no direct baseline comparison is available, the results demonstrate that GitFarm efficiently supports this class of workflows with minimal overhead. This use case also illustrates GitFarm’s ability to support output-dependent command chaining within a single execution session.

\subsubsection{Case Study 3: Read-only Inspection of up-to-date Code Ownership Files}

Many automated systems require frequent access to repository metadata and file contents, but do not modify repository state. One such production workload determines code ownership by scanning ownership files across hundreds of thousands of directories in Uber’s monorepos (Go monorepo has more than 350K subdirectories). 

\paragraph{Setup.}
The workload is deployed on a production fleet of six hosts and issues read-only Git operations at an average rate of approximately 100 requests per second. Each host maintains a local Git checkout for every supported monorepo (Go, Java, Web, Python, iOS, and Android), which are periodically synchronized using a cron job that runs every 15 minutes.

\paragraph{Metrics.}
We evaluate the impact of GitFarm in terms of overall resource utilization, with an emphasis on CPU, memory and disk space usage.

\begin{figure}[htbp]
    \centering
    \includegraphics[width=1\linewidth]{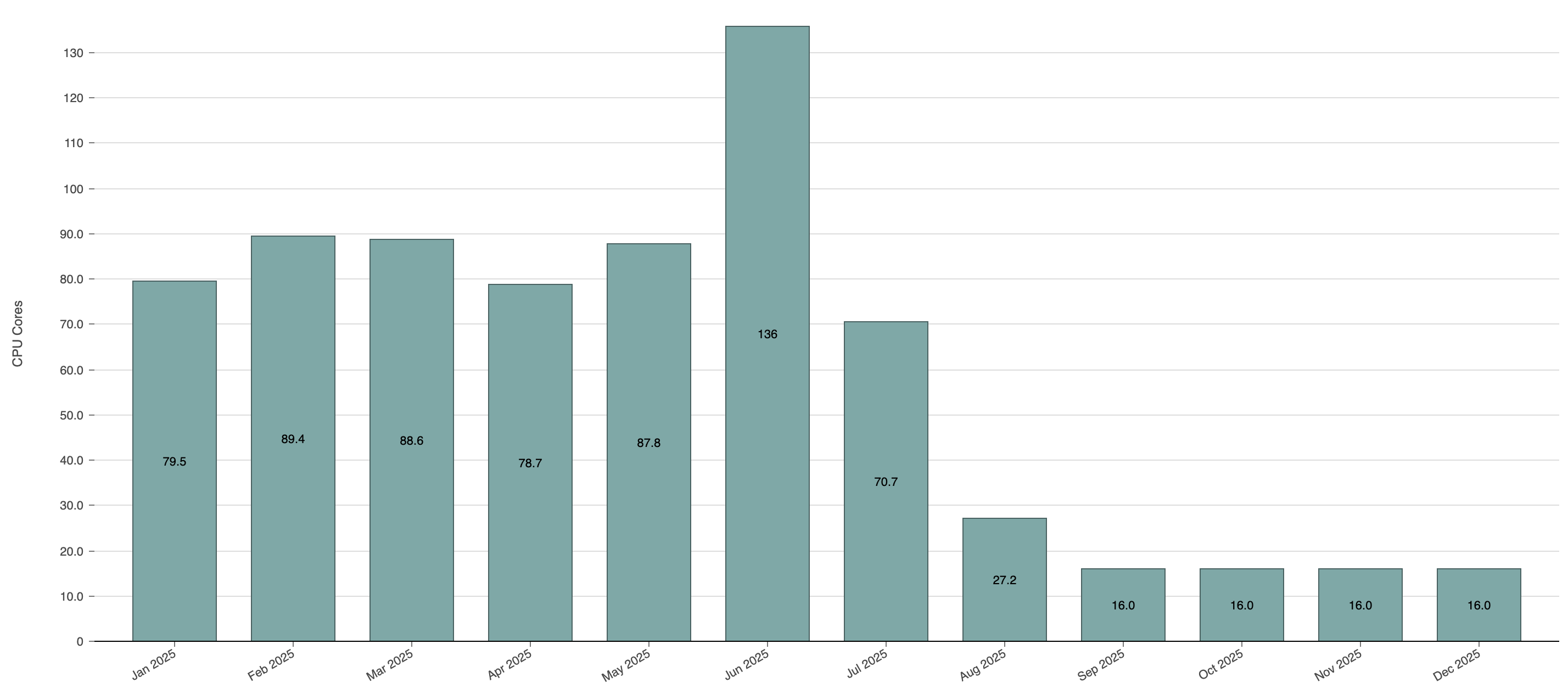}
    \caption{CPU Cores used before and after eliminating Local Checkouts}
    \Description{CPU Cores used before and after eliminating Local Checkouts}
    \label{fig:code-metadata-cpu-usage}
\end{figure}

\begin{figure}[htbp]
    \centering
    \includegraphics[width=1\linewidth]{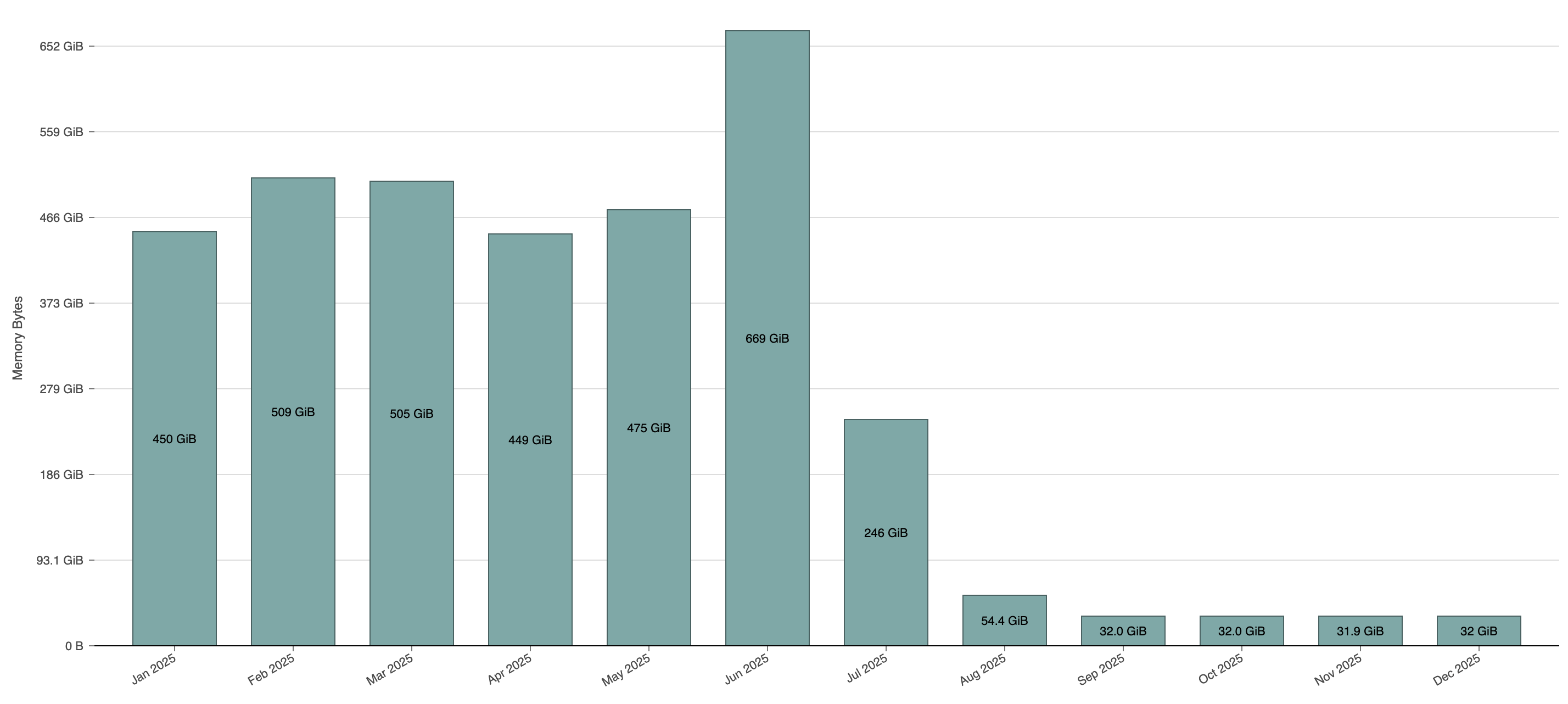}
    \caption{Memory used before and after eliminating Local Checkouts}
    \Description{Memory used before and after eliminating Local Checkouts}
    \label{fig:code-metadata-mem-usage}
\end{figure}

\paragraph{Resource Utilization.}
After the service was able to remove the full repository checkouts of the monorepos, per-node resource allocations were significantly reduced. CPU cores per node decreased from 12 to 2, memory from 48 GB to 4 GB, and disk capacity from 64 GB to 2 GB, sufficient for service-related files only. Figures \ref{fig:code-metadata-cpu-usage} and \ref{fig:code-metadata-mem-usage} show the resulting fleet-wide average resource utilization before and after the transition in August 2025. Aggregate CPU consumption dropped from an average of 90.1 cores in the 7 months leading up to August 2025 to 16 cores after the transition was completed (an 82\% reduction). Similarly, memory usage decreased from an average of about 472 GB to 32 GB (a 93\% reduction). As a result, the cost of running this service scaled down proportionally.

\paragraph{Startup Time.}
When the service maintained local checkouts, each node performed local repository cloning and synchronization during initialization, resulting in startup times of approximately 15–20 minutes per node, due to the large monorepos. After eliminating local checkouts, nodes no longer need to perform the large clone, reducing the time to start up to under one minute, enabling significantly faster scaling, deployment, and rollbacks.

\paragraph{Discussion.}
These results show that centralizing repository access through GitFarm can substantially improve resource efficiency and operability for read-heavy automation services. Removing local checkouts eliminates redundant synchronization and large on-disk state, reducing steady-state resource usage and enabling faster service startup.

\section{Related Work}

This section surveys prior efforts related to scaling version control systems, improving Git performance, and supporting Git-centric automation workloads. These approaches address different aspects of the problem space but do not directly target centralized, service-oriented execution of Git operations for automation.

\subsection{Scaling Version Control Systems}

Large-scale monorepos have motivated several organizations to re-architect their version control systems. Google’s Piper \cite{potvin2016googlemonorepo} replaced Git with a centralized source control backend designed to scale to billions of files and tens of thousands of developers. Similarly, Meta’s Mononoke \cite{facebook2019mononoke} and Sapling \cite{facebook2022sapling} extended Mercurial to support extremely large repositories with fast branching and history traversal.

These systems demonstrate that scaling monolithic codebases often requires rethinking traditional distributed version control models. However, they are proprietary, tightly integrated with internal tooling, and not designed to expose Git-compatible interfaces for general-purpose automation workloads.

\subsection{Scaling Git Hosting Infrastructure}

Several efforts focus on scaling Git server infrastructure while preserving Git’s distributed semantics. GitHub’s Spokes \cite{github2016spokes} and GitLab’s Gitaly Cluster \cite{gitlab2023gitalycluster} horizontally scale Git hosting by replicating repositories across backend nodes and distributing client traffic. While these designs improve aggregate throughput, they still require strong consistency among replicas to preserve object integrity, reference correctness, and garbage collection semantics.

As a result, server-side operations such as object enumeration, packfile generation, and data streaming must be performed independently for each client request \cite{chacon2024progitpackfiles,git2024packobjects}, limiting scalability under high concurrency.

\subsection{Client-Side Git Optimizations}

Prior work has explored improving Git performance by optimizing client-side repository access. Microsoft’s GVFS and its successor, Scalar \cite{microsoft2020scalar}, introduced virtualized working trees and background fetching to reduce clone and checkout costs for large repositories. The Git community has also introduced shallow clones and partial clones \cite{git2024partialclone} to reduce data transfer during initial checkout.

While these techniques improve individual client performance, they do not eliminate server-side computation during clone and fetch operations, and often impose functional limitations. Shallow history restricts operations such as \textit{merge-base} and \textit{bisect}, and partial clones frequently break tooling assumptions in large monorepos \cite{github2021monorepomaintenance}.

\subsection{CI-Based Caching and Repository Reuse}

CI platforms commonly employ repository caching mechanisms to reduce clone latency. Jenkins uses persistent workspaces and SCM plugins to reuse local repositories across jobs \cite{jenkinsgitplugin2023reference,jenkins2024pipelinescmstep}, while Buildkite agents maintain local mirror caches to avoid redundant network transfers \cite{buildkite2023caching,buildkite2020issue734}.

These optimizations operate at the agent level and remain tightly coupled to CI job execution. Repository state is scoped to individual agents, requires manual maintenance, and does not eliminate redundant synchronization across independent automated systems. As a result, CI-based caching improves per-job efficiency but does not address the broader challenge of consolidating Git access across services.

\subsection{Sandboxed Remote Execution and Service Abstractions}

GitFarm’s execution model draws inspiration from distributed build and remote execution systems such as Google’s Remote Execution API \cite{google2020reapi}, as well as secure container-based isolation environments like gVisor \cite{gvisor2018hotcloud}. These systems demonstrate the benefits of isolating workloads, pooling warm execution environments, and minimizing startup latency through reuse.

Cloud-native management services such as AWS Systems Manager (SSM) \cite{aws-ssm} further illustrate the feasibility of identity-aware remote command execution at scale. However, these systems operate at the infrastructure management layer and lack repository-aware execution semantics, persistent workspace state, and multi-command session guarantees required for complex Git workflows.

GitFarm applies these remote execution principles specifically to version control, executing Git commands within pre-initialized sandboxes backed by pooled, pre-synchronized repositories. This repository-centric execution model enables efficient, multi-step Git automation while preserving native Git semantics.

\section{Limitations and Future Work}

While the current GitFarm system is robust and effective, several enhancements remain to further improve efficiency and flexibility across a broader range of workloads.

\subsubsection{Streaming Git Operation Output}
At present, GitFarm returns command output as monolithic stdout/stderr payloads. Supporting streaming output would allow large result sets, such as file contents, to be transmitted incrementally rather than returned as monolithic strings, reducing memory pressure and improving efficiency for large responses.

\subsubsection{Support for Alternative Repository Workspaces}
GitFarm currently operates exclusively on full repository checkouts, which can be inefficient for large repositories. Supporting alternative workspace models, such as sparse checkouts, would enable faster and more resource-efficient operations for workloads that do not require full working trees.

\subsubsection{Long-Lived Sessions}
GitFarm’s sessions are optimized for short-lived execution and are capped at 5 minutes. However, certain workflows benefit from substantially longer-lived execution contexts. For example, services that continuously observe changes on a long-lived branch (e.g., \texttt{main}) to monitor file updates or compute derived metadata could maintain a persistent execution session over extended periods. Retaining a warm repository state across such sessions would reduce repeated initialization overhead and enable more efficient, low-latency processing of successive change events.

\subsection{Planned Integrations}

In addition to the evaluated use cases, GitFarm is designed to support a broader set of Git-intensive automation workloads planned for future integration.

\subsubsection{Mirroring Changes Across Git Upstreams}
Backup and replication services will leverage GitFarm to continuously monitor repository push events and mirror new commits from a primary Git upstream to a secondary upstream. By centralizing repository state, these services will avoid maintaining local checkouts, while ensuring consistent and up-to-date backups across thousands of repositories.

\subsubsection{Merge to Trunk (or Main)}
SubmitQueue (Uber’s merge queue system) \cite{ananthanarayanan2019keeping} will leverage GitFarm to validate incoming changes by applying them onto the latest trunk (e.g. \texttt{main}) branch within sandboxed environments. Validated commits will be pushed under a temporary reference for deferred merging. Once CI validation completes, SubmitQueue will invoke GitFarm to cherry-pick commits from this reference, perform the final merge, and push to trunk. This will enable efficient and isolated merge processing without requiring local repository state.

\section{Conclusion}

GitFarm represents a major step forward in how Git operations are executed at Uber’s scale. By providing Git as a Service, we have removed the burden of repository management from individual systems, offering a faster, more secure, and scalable alternative to traditional workflows. As the platform evolves with features like streaming output, flexible workspace types, and extended sessions, it will continue to enable seamless, high-performance Git operations that keep pace with Uber’s growing monorepos and engineering ecosystem.

\bibliographystyle{ACM-Reference-Format}
\bibliography{references}
\end{document}